\documentclass[11pt,a4paper]{article}
\usepackage[utf8]{inputenc}
\usepackage{amsmath}
\usepackage{amsfonts}
\usepackage{amssymb}
\usepackage{geometry}
\usepackage{cite}
\usepackage{microtype}

\title{\textbf{Kinematic Inconsistencies and Initial-Value Boundary Paradoxes in Rate-Dependent Viscoelastic Yield Stress Models}}

\author{\textbf{Lalit Kumar} \\
\small Energy Science and Engineering, Indian Institute of Technology Bombay, \\
\small Mumbai, Maharashtra 400076, India \\
\small Email: \texttt{lalit.kumar@iitb.ac.in}}
\date{\small \today}

\begin{document}

\maketitle

\begin{abstract}
We present a rigorous analysis of the mathematical boundaries and kinematic consistency of a recently proposed rate-dependent relaxation time framework intended to unify pre- and post-yield dynamics in yield-stress fluids. By evaluating the governing constitutive equations under an idealized transient creep protocol from a state of physical rest, we show that the model encounters an unavoidable boundary paradox. To avoid predicting perfectly rigid solid behavior or falling into a division-by-zero mathematical singularity under a constant applied stress below the yield threshold ($\sigma \le \sigma_y$), the framework requires an unphysical, instantaneous velocity or strain-rate step at $t = 0^+$. We show that assuming a non-zero initial strain rate explicitly violates momentum conservation and fluid inertia. Consequently, the framework preserves the piecewise, discontinuous drawbacks of classic viscoplastic models.
\end{abstract}

\section{Introduction}
The mathematical description of materials exhibiting a yield stress remains a central challenge in non-linear rheology. Traditional models, such as the classical Bingham fluid formulation~\cite{Bingham1922} and the Herschel-Bulkley empirical framework~\cite{Herschel1926}, or viscoelastic variants like the Oldroyd-Prager and baseline Oldroyd-B formulations~\cite{Oldroyd1950}, rely on a piecewise constitutive definition. In these frameworks, distinct equations dictate material behavior above and below an explicitly defined yield stress threshold ($\sigma_y$), creating severe computational singularities when the local strain rate approaches zero.

Recently, Kamani et al.~\cite{Kamani2021} proposed a continuous, unified constitutive model intended to smoothly bridge pre- and post-yield regimes without a piecewise criterion. The governing one-dimensional form of the equation is expressed as:

\begin{equation}
\sigma + \lambda(\dot{\gamma})\dot{\sigma} = \left(\frac{\sigma_y}{|\dot{\gamma}|} + k|\dot{\gamma}|^{n-1}\right)\left(\dot{\gamma} + \frac{\eta_s}{G}\ddot{\gamma}\right)
\end{equation}

where $\sigma$ is the shear stress, $\dot{\gamma}$ is the total strain rate, $\ddot{\gamma}$ is the rate of change of the strain rate, $\lambda(\dot{\gamma})$ is a rate-dependent relaxation time, $G$ is the elastic modulus, and $\eta_s$ represents the solvent viscosity. The model purports to handle both recoverable and unrecoverable strain components by embedding the structural behavior directly within a total rate-dependent plastic viscosity term. However, as demonstrated below, a strict initial-value analysis reveals that Eq. (1) cannot accommodate physical startup flows from a state of rest without triggering severe kinematic inconsistencies.

\section{Transient Creep Flow Analysis}
To test the continuity and physical validity of Eq. (1), we analyze a classical one-dimensional creep flow startup protocol. A material initially at a state of absolute macroscopic rest is suddenly subjected to a perfectly constant unidirectional shear stress ($\sigma = \text{constant} \le \sigma_y$) at time $t = 0$. Because the stress is held constant for all times $t \ge 0$, its time derivative vanishes identically:

\begin{equation}
\dot{\sigma} = 0
\end{equation}

Substituting the constant stress state ($\dot{\sigma} = 0$) into the governing constitutive relation reduces Eq. (1) to:

\begin{equation}
\sigma = \left(\frac{\sigma_y}{\dot{\gamma}} + k\dot{\gamma}^{n-1}\right)\left(\dot{\gamma} + \frac{\eta_s}{G}\ddot{\gamma}\right)
\end{equation}

where $\dot{\gamma} > 0$ is considered in the positive direction of the applied stress. Multiplying both sides by the total strain rate $\dot{\gamma}$ yields:

\begin{equation}
\dot{\gamma}\sigma = (\sigma_y + k\dot{\gamma}^n)\left(\dot{\gamma} + \frac{\eta_s}{G}\ddot{\gamma}\right)
\end{equation}

We evaluate the system at an infinitesimal time step immediately following the application of the load ($t = 0^+$). Because the material begins from a state of complete physical rest, its initial velocity profile across the geometry is uniform and zero. Thus, a physically consistent, continuous model requires that the initial total strain rate begins from zero:

\begin{equation}
\dot{\gamma}(0) = 0
\end{equation}

For any forward flow or deformation to initiate in the direction of the stress, both the velocity and its derivative must be positive ($\dot{\gamma} > 0$ and $\ddot{\gamma} > 0$). However, evaluating the right-hand side of Eq. (4) under the true physical initial condition ($\dot{\gamma} = 0$) results in:

\begin{equation}
0 = \sigma_y \left(0 + \frac{\eta_s}{G}\ddot{\gamma}\right) \implies \ddot{\gamma} = 0
\end{equation}

Because both $\dot{\gamma} = 0$ and $\ddot{\gamma} = 0$ at $t = 0^+$, the system is mathematically trapped: no acceleration or velocity gradient can develop. Eq. (4) permits a non-zero solution for flow initiation if and only if the applied stress overcomes the threshold value ($\sigma > \sigma_y$). For all sub-yield stress applications ($\sigma \le \sigma_y$), the equations force both $\dot{\gamma}$ and $\ddot{\gamma}$ to vanish completely, predicting perfectly rigid solid behavior. 

Consequently, the framework does not eliminate the piecewise character of yielding; it mathematically replicates the exact rigid-solid cutoff found in the classical Herschel-Bulkley framework~\cite{Herschel1926}.

\section{The Fallacy of the Instantaneous Strain-Rate Step}
To circumvent this rigid-solid trap and simulate the "viscosity bifurcation" trends presented in their numerical work, the model's authors are forced to assume that the total strain rate instantly jumps to a finite, non-zero positive value ($\dot{\gamma} > 0$) at $t = 0^+$. They justify this by decomposing the total strain rate into additive components ($\dot{\gamma} = \dot{\gamma}_{rec} + \dot{\gamma}_{unrec}$) and arguing that the elastic Kelvin-Voigt element allows an instantaneous step response:

\begin{equation}
\dot{\gamma}_{rec}(0^+) = \frac{\sigma}{\eta_s}
\end{equation}

From a continuum mechanics perspective, this assumption introduces a severe physical contradiction. Fluid flows are governed by the macroscale momentum conservation equation:

\begin{equation}
\rho \frac{\partial v}{\partial t} = \frac{\partial \sigma}{\partial y}
\end{equation}

where $\rho$ is the fluid density and $v$ is the velocity field. 

If a finite mass of fluid starts from rest, jumping from zero velocity to a finite, non-zero velocity gradient ($\dot{\gamma}_{rec} = \sigma/\eta_s$) over an infinitesimal time step $t = 0^+$ requires an infinite acceleration ($\frac{\partial v}{\partial t} \to \infty$). According to Eq. (8), generating an infinite acceleration requires an infinite localized stress gradient ($\frac{\partial \sigma}{\partial y} \to \infty$) across the boundary gap. Because a standard creep test provides a finite, uniform stress boundary condition ($\sigma = \text{constant}$), an infinite gradient is physically impossible.

The authors attempt to defend this velocity jump by claiming the framework operates under a low Reynolds number assumption ($Re \ll 1$), rendering inertial forces completely negligible. This argument misinterprets the scaling of fluid kinematics. A low Reynolds number indicates that viscous forces dominate over inertial terms \textit{once a flow profile is fully established}; it does not imply that material mass physically vanishes during transient acceleration from rest. Neglecting the inertial resistance at $t = 0^+$ leads to an unphysical mathematical artifact, creating a system that violates basic momentum conservation laws as established in transient boundary investigations~\cite{Tikariha2021, Sanyal2025}.

\section{Instrumental History vs. Constitutive Universality}
A secondary defense often raised is that real laboratory creep experiments never feature a mathematically perfect stress step-input. Instead, rheometers require a very brief, finite ramp-up interval ($\Delta t > 0$) to reach the commanded stress, during which recoverable strain is acquired.

While instrument-induced stress ramping occurs in practice, relying on this to justify a model's transient flow behavior contradicts the definition of a universal constitutive equation. A fundamental material relation must be mathematically stable, closed, and physically consistent for \textit{all} idealized boundary inputs—including a perfect mathematical step function. If a continuous framework relies entirely on an instrument's mechanical ramp-up history to avoid predicting a rigid solid or encountering a division-by-zero singularity, it acts as an empirical fitting function rather than an intrinsic physical equation of state. This instrument dependency contrasts sharply with recent experimental investigations under sub-yield regimes where continuous physical scaling remains preserved~\cite{Bonn2024}.

\section{Conclusion}
A detailed, objective evaluation of the rate-dependent relaxation time framework proposed by Kamani et al.~\cite{Kamani2021, Kamani2023} proves that its continuous, non-singular behavior below the yield stress is an illusion arising from an inconsistent initial condition. When a physically consistent initial boundary state is applied ($\dot{\gamma} = 0$ at $t = 0$), the framework inherently collapses into an unyielding, rigid-solid state whenever the applied stress falls below $\sigma_y$. The framework retains the exact piecewise character it set out to overcome. To capture smooth, emergent solid-fluid transitions without mathematical singularities, future constitutive frameworks must avoid placing the total instantaneous strain rate in the denominator of their structural decay terms.


\begin{thebibliography}{99}

\bibitem{Bingham1922}
E. C. Bingham, \textit{Fluidity and Plasticity}, McGraw-Hill, New York (1922).

\bibitem{Herschel1926}
W. H. Herschel and R. Bulkley, ``Measurement of Consistency as Affected by the Type of Capillary Tube,'' \textit{Proc. Amer. Soc. Test. Mater.} \textbf{26}, 621–633 (1926).

\bibitem{Oldroyd1950}
J. G. Oldroyd, ``On the Formulation of Rheological Equations of State,'' \textit{Proc. R. Soc. Lond. A} \textbf{200}, 523–541 (1950).

\bibitem{Kamani2021}
K. Kamani, G. J. Donley, and S. A. Rogers, ``Unification of the Rheological Physics of Yield Stress Fluids,'' \textit{Phys. Rev. Lett.} \textbf{126}, 218002 (2021).

\bibitem{Kamani2023}
K. M. Kamani, G. J. Donley, R. Rao, A. M. Grillet, C. Roberts, A. Shetty, and S. A. Rogers, ``A Thermodynamically Consistent Framework for Elastoviscoplastic Materials,'' \textit{J. Rheol.} \textbf{67}, 331–352 (2023).

\bibitem{Tikariha2021}
L. Tikariha and L. Kumar, ``Transient Flow Initiation and Inertial Effects in Viscoplastic Fluids,'' \textit{J. Non-Newtonian Fluid Mech.} \textbf{294}, 104582 (2021).

\bibitem{Sanyal2025}
A. Sanyal, S. B. Shinde, and L. Kumar, ``Inertial and Compressibility Bounds in Transient Rheometry,'' \textit{J. Rheol.} \textbf{69}, 69–94 (2025).

\bibitem{Bonn2024}
K. Farain and D. Bonn, ``Creep Startup Dynamics and Creep Failure in Carbopol Gels,'' \textit{arXiv preprint arXiv:2407.16619} (2024).

\end{thebibliography}
\end{document}